\documentclass[a4paper,12pt]{article}

\usepackage{graphicx}
\usepackage{amsmath,amssymb}

\begin{document}

\title{\bf 3D Calculation with Compressible LES\\
for Sound Vibration of Ocarina}
\author{Taizo KOBAYASHI$^\dagger$, Toshiya TAKAMI$^\dagger$,
Masataka MIYAMOTO$^\ddagger$,\\
Kin'ya TAKAHASHI$^\ddagger$, Akira NISHIDA$^\dagger$,
and Mutsumi AOYAGI$^\dagger$\\
\\
\it $^\dagger$Research Institute for Information Technology,
Kyushu University,\\
\it 6-10-1 Hakozaki, Higashi-ku, Fukuoka 812-8581, JAPAN\\
\it $^\ddagger$Physics Laboratories, Kyushu Institute of Technology,\\
\it Kawazu 680-4, Iizuka 820-8502, JAPAN}
\date{}
\maketitle

\abstract{Sounding mechanism is numerically analyzed to elucidate
physical processes in air-reed instruments.  As an example, compressible
large-eddy simulations (LES) are performed on both two and three
dimensional ocarina.  Since, among various acoustic instruments, ocarina
is known as a combined system consisting of an edge-tone and a Helmholtz
resonator, our analysis is mainly devoted to the resonant dynamics in
the cavity.  We focused on oscillation frequencies when we blow the
instruments with various velocities.}

\section{Introduction}

Elucidation of acoustical mechanism of air-reed instruments is a long
standing problem in the field of musical acoustics\cite{Fletcher}.  The
major difficulty of numerical calculations of an air-reed instrument is
in strong and complex interactions between sound field and air flow
dynamics\cite{Coltman}, which is hardly reproduced by hybrid methods
\cite{HPCAsia} normally used for analysis of aero-acoustic
noises\cite{LES-book,Noise}.

There are two types of air-reed instruments, which are different in
acoustics mechanism: in the group of flute, recorder, organ pipe, the
pitch of an excited sound is determined by the length of an air column,
i.e., resonance of the air column; on the other hand, it is considered
that sound of an ocarina is produced by Helmholtz resonance where the
pitch is determined by resonance of the entire cavity and the placement
of each hole on an ocarina is almost irrelevant.  It should be noted
that the Helmholtz resonance is based on an elastic property of the air,
not the sound propagation.  This is another reason that the usual hybrid
model consisting of fluid mechanics and sound propagation cannot
reproduce the oscillating dynamics in an ocarina.  Thus, when we study
the acoustic mechanism of the ocarina, the whole calculation of a
compressible fluid mechanics for the air-reed and the resonator is
essentially required.


Taking an ocarina as a model system, we investigate how the Helmholtz
resonator is excited by the edge tone\cite{Brown} created by a jet flow
collided to an edge of aperture of the cavity.  The ocarina is a
relevant model for our purpose, since it creates a clear tone and is
enough small in size to calculate with present computational resources.

In addition to the sound propagation, the oscillation in an ocarina is
described by an elastic dynamics of air.  In the present calculation,
the LES solver {\tt coodles} in OpenFOAM 1.5 is used to directly solve
the compressible Navier Stokes equation in order that both the radiated
sound and flow dynamics are simultaneously reproduced.  This paper is
organized as follows.  In Section~\ref{sec:2d}, by the use of two
dimensional ocarina model, we explore suitable playing conditions with
changing the velocity such that a well sustained sound vibration is
excited in the cavity.  In Section~\ref{sec:3d}, a reproduced sound by
three dimensional ocarina model is analyzed.  Based on these analysis, a
conclusion is given in Section~\ref{sec:conclusion}.

\section{Two dimensional ocarina}
\label{sec:2d}

\begin{figure}
 \centering{\includegraphics[width=10cm]{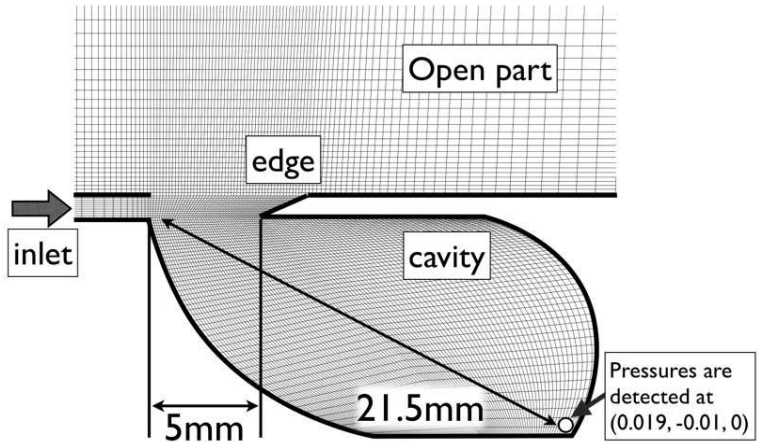}}
 \caption{\small Geometry of the 2D ocarina.  There are no
 tone holes other than the aperture in the upper-left corner of the cavity.
 A jet is blown from the left duct.  The area of the cavity region
 is $1.635$ cm$^2$.
 }
 \label{fig:2dMesh}
\end{figure}

At first, a two-dimensional air-reed instrument is configured.
The geometry studied in this section is shown in Fig.\ref{fig:2dMesh}.
The aperture of this instrument is $0.5$ cm, and the area of the cavity
is $1.635$ cm$^2$.  The maximum length in the cavity is $21.5$ mm.

Numerical calculations are performed by OpenFOAM-1.5.
In order to solve in two-dimension,
front and back planes in z-direction are set to {\tt empty} type.
The following boundary conditions are introduced:
the fluid velocity and the pressure gradient are set to zero on walls
({\tt fixedValue} and {\tt zeroGradient} are used);
{\tt inletOutlet} walls with $U=(0,0,0)$ m/s and $p=10^5$ Pa
are introduced at boundaries of the open part.

We are interested in relation between the pitch of sound vibration and
the jet velocity.  The excitation of over-tone will be prohibited if the
sound vibration in ocarina mainly originates from the Helmholtz
resonance.  Then, the pitch of fundamental only depends on the volume
of the cavity but irrespective to its shape.

\subsection{Frequency of Helmholtz resonance}
\label{sec:2dHelmholtz}

The resonance frequency of our two-dimensional model is estimated.  We
approximate that the effective length $L$ of open end correction is in
proportion to the length of aperture $l$ and the proportional constant
is $\alpha$.  Then the effective mass $m$ and the spring constant $K$
for Helmholtz resonator are given by
\begin{equation}
 \begin{split}
  m &= \rho l L \cong \rho \alpha l^2 \\
  K &= \rho l^2c^2/s,
 \end{split}
\end{equation}
where $\rho$, $c$, and $s$ are the density of the air, the sound
velocity, and the size of the cavity, respectively.

The resonant frequency $f_0$ can be written as follows:
\begin{equation}
 f_0 = \frac{1}{2\pi}\sqrt{\frac{K}{m}}.
\end{equation}
Thus, the frequency of the two-dimensional Helmholtz resonance is
\begin{equation}
 f_{2D} \approx \frac{c}{2\pi \sqrt{\alpha s}}.
\label{eq:f2d}
\end{equation}

Next, let's estimate the $f_{2D}$ for our 2D model.
Parameters of the model are
\begin{equation}
\begin{split}
 s &= 1.6351 \times 10^{-4} \\
 l &= 5.0 \times 10^{-3} \\
 c &= 340.
\end{split} 
\end{equation}
By the use of the above parameters, the value of Eq.~(\ref{eq:f2d}) is
estimated to
\begin{equation}
 f_{2D} \approx 4.232 \times 10^{3} \times \alpha ^{-\frac{1}{2}}.
  \label{eq:estimate}
\end{equation}

\subsection{Sounds generated in the 2D model}

\begin{figure}
 \centering{
 \includegraphics[width=6cm]{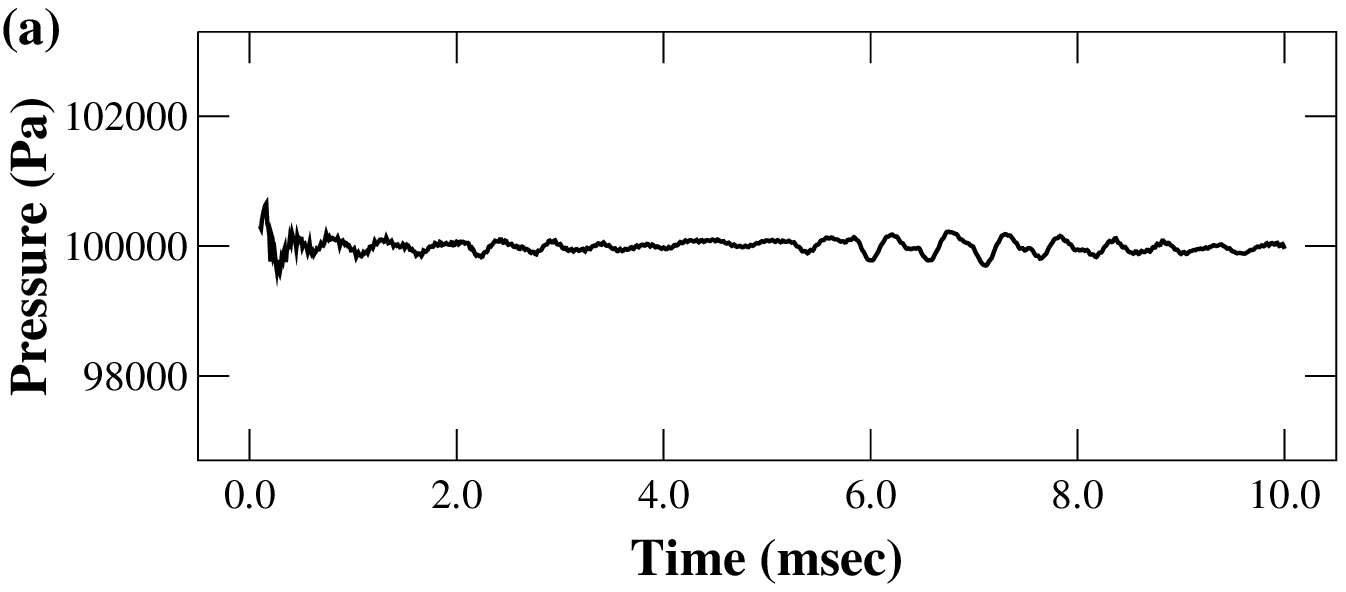}\hspace{5mm}
 \includegraphics[width=6cm]{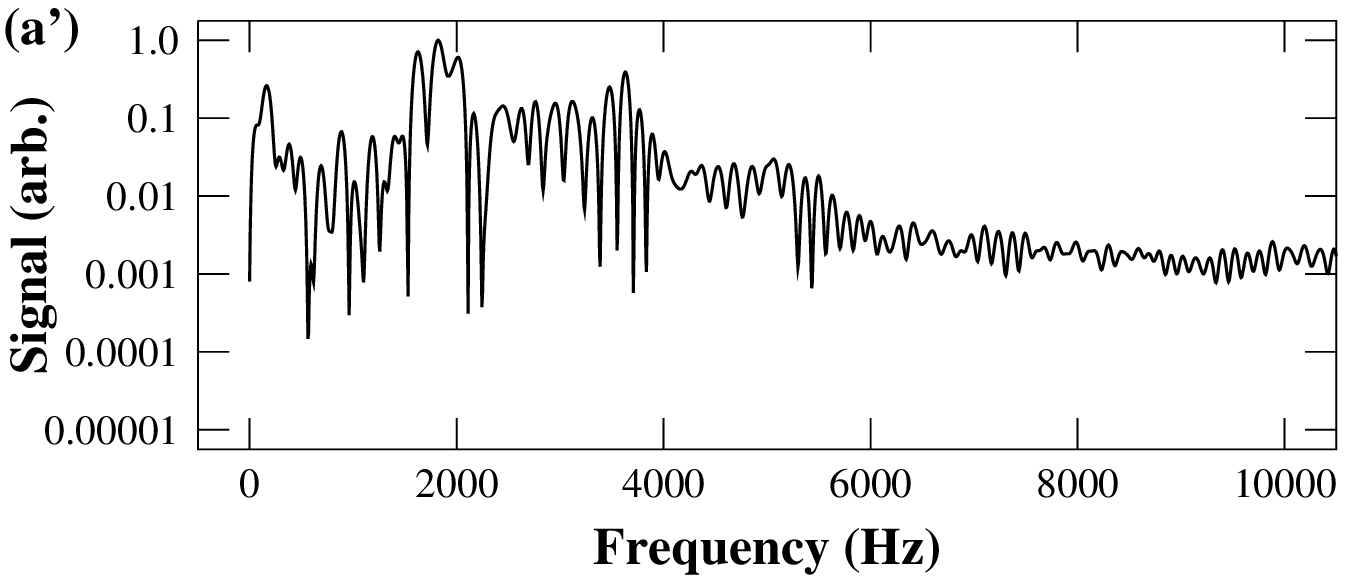}\\
 \includegraphics[width=6cm]{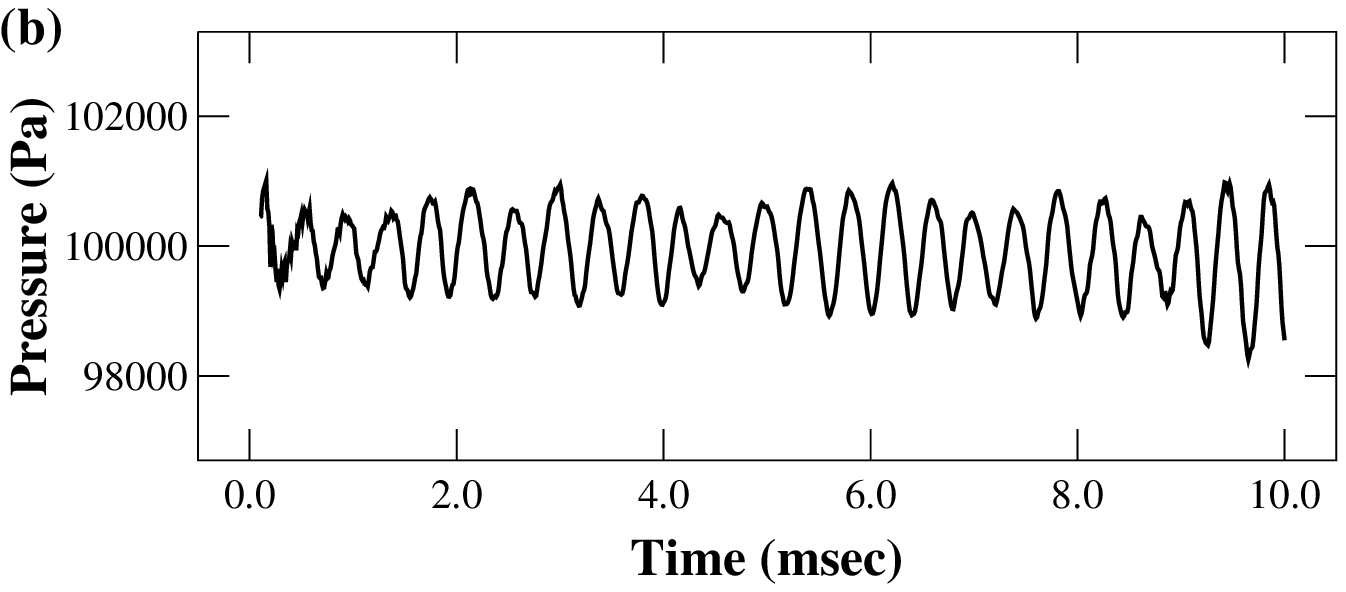}\hspace{5mm}
 \includegraphics[width=6cm]{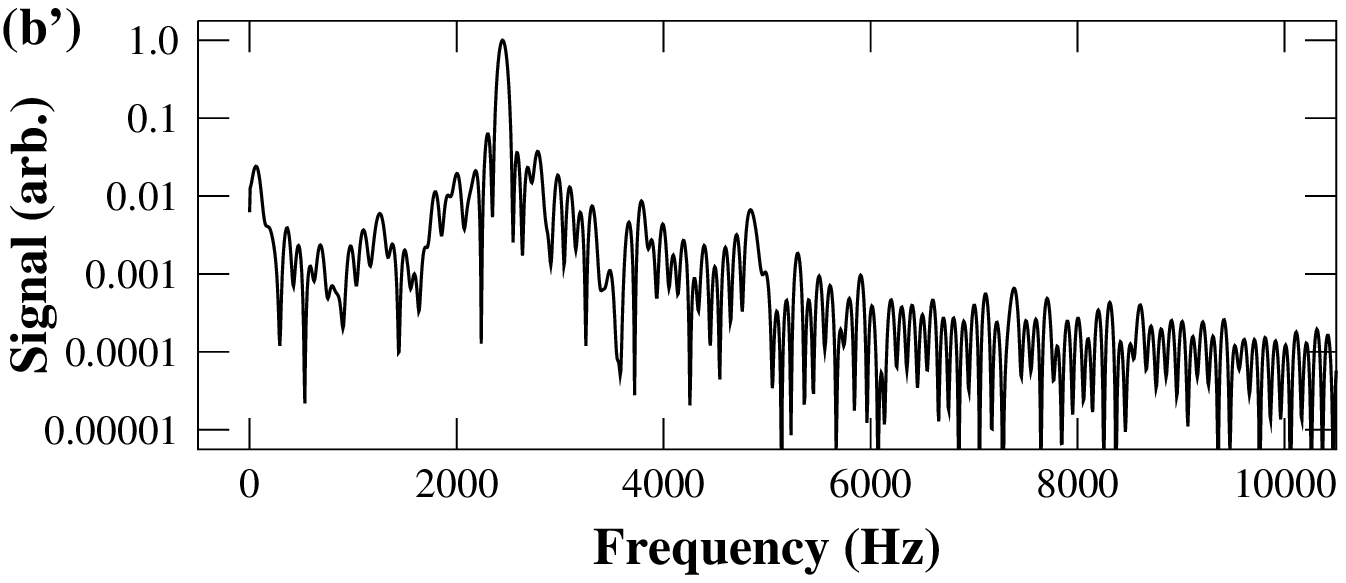}\\
 \includegraphics[width=6cm]{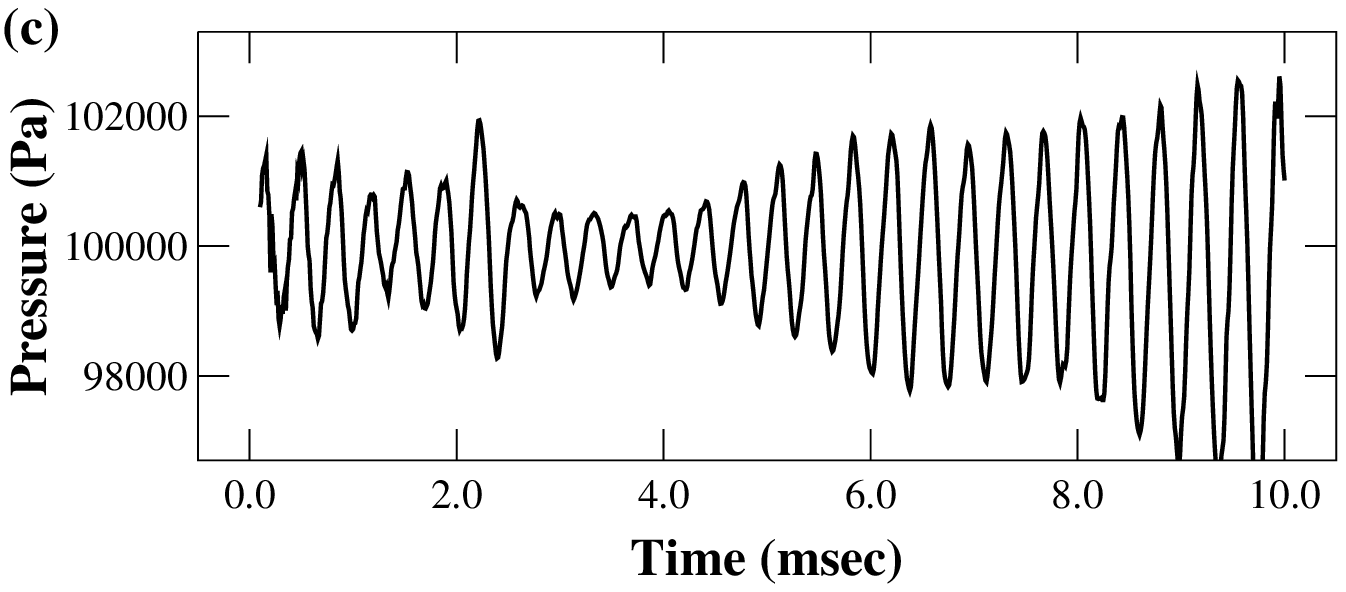}\hspace{5mm}
 \includegraphics[width=6cm]{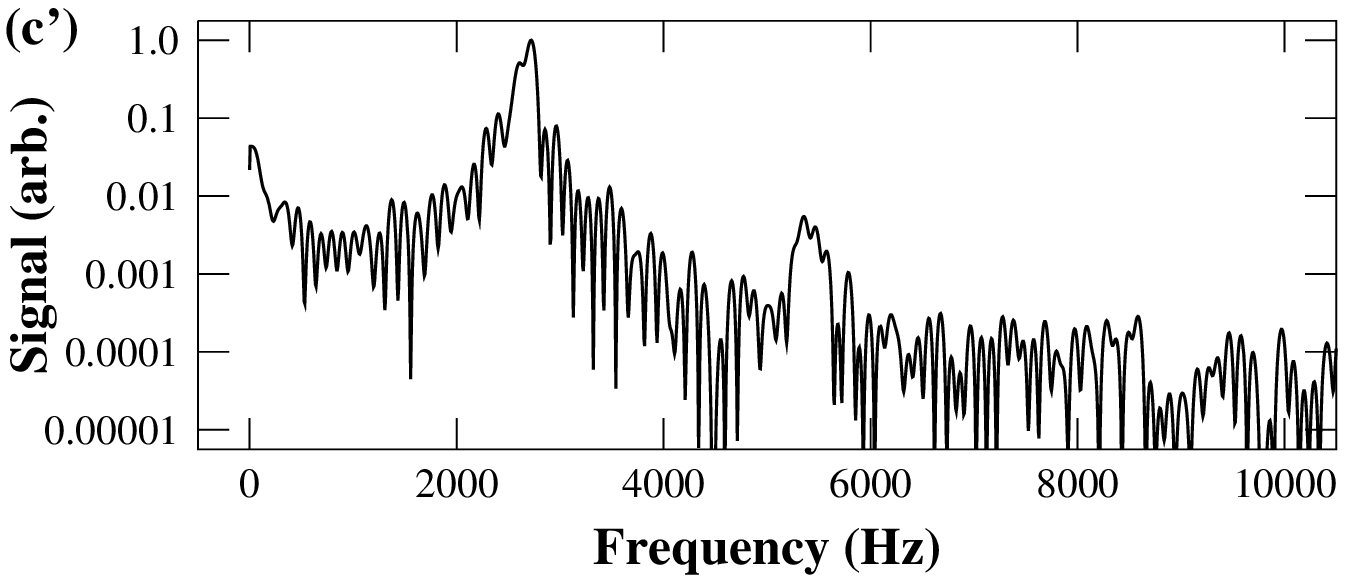}
 }
 \caption{\small Pressure in 2D ocarina with blow velocities:
 (a) 20 m/s; (b) 30 m/s; (c) 40 m/s.
 The corresponding normalized spectrum is shown in (a'), (b'), and (c').}
 \label{fig:2dP}
\end{figure}

Numerically obtained pressure values and their normalized spectrum
are shown in Fig.\ref{fig:2dP}.
While the oscillation is very small
and unstable for the blown velocity $20$ m/s, 
the oscillations are sufficiently grown for velocities
$30$ m/s and $40$ m/s.
In the figures of the normalized spectrum,
almost no higher harmonics are observed in (b') and (c').
The single oscillation without harmonics is one of the
properties of the Helmholtz resonance.

\begin{figure}
 \centering{\includegraphics[width=12cm]{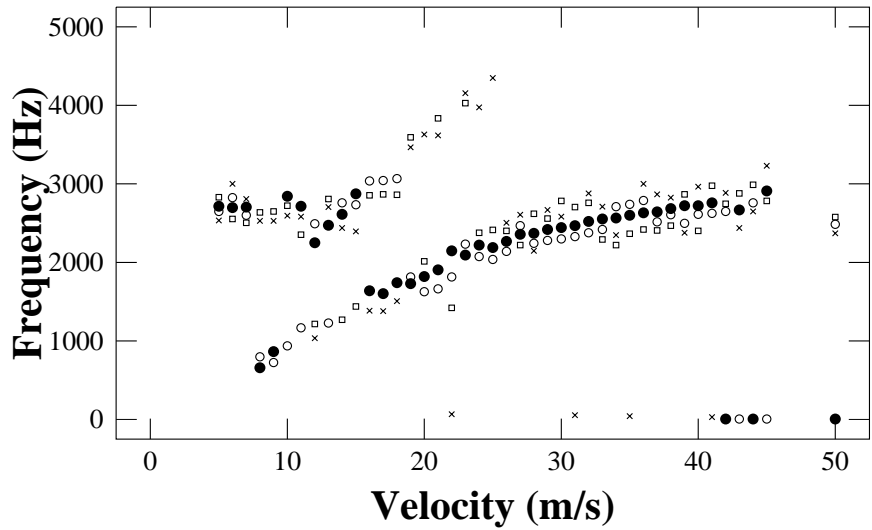}}
 \caption{\small Peak frequencies of the 2D ocarina for various blow
 velocities.  The first, second, third, and fourth peaks in each
 velocity are shown in filled circles ($\bullet$), open circles
 ($\circ$), small squares ({\tiny $\square$}), and crosses ($\times$),
 respectively.}
 \label{fig:peaks}
\end{figure}

The frequency for the most prominent peak in each Fourier
transformed data of the pressure are shown in Fig.\ref{fig:peaks}.
As the velocity becomes larger, the frequency slightly increases.
This shows a broad resonance which is often observed in ocarina, and
the frequency as the Helmholtz resonance is about $2.5\times10^3$ Hz.
It is very difficult to obtain only from theoretical considerations
with analogies from those simple Helmholtz instruments found in
textbooks\cite{Fletcher}.
If we use this resonant frequency $2.5\times10^3$ Hz for the formula
in Section \ref{sec:2dHelmholtz}, the effective value of the open end
correction for this model is determined as $\alpha\approx2.87$
which is large in comparison with $\alpha\approx 16/3\pi$ in three-dimension.

The open end correction can
be affected by the blow velocity (jet) on the aperture.
One of the reason for the disagreement is that
this effect is not considered in the estimation Eq.(\ref{eq:estimate}).

\section{Three dimensional ocarina}
\label{sec:3d}

\begin{figure}
\centering{
\includegraphics[width=10cm]{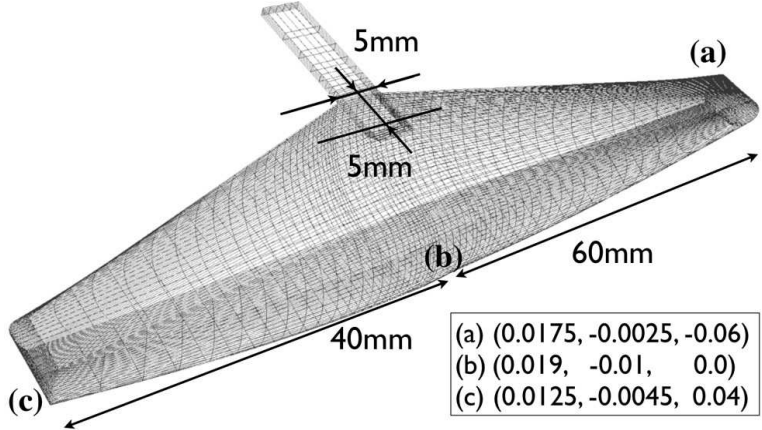}
}
\caption{\small The whole geometry of the 3D ocarina model.
In this figure, only thinned mesh points are shown.
(a), (b), and (c) represent pressure observation points.
}
\label{fig:3dMesh}
\end{figure}

Three dimensional ocarina model investigated here
is shown in Fig.\ref{fig:3dMesh}.
The numbers of mesh points for numerical calculations
are tabulated in Table \ref{tbl:mesh}.
A flow is added from the inlet duct and emitted as a jet
at a $5 \times 5$ mm$^2$ aperture to hit an edge.
An open-box with $100 \times 50 \times 100$ mm$^3$
is introduced above the aperture.
The boundary conditions are the same as the two-dimensional case.
In this section, the velocity of the jet is fixed to 10 m/s.

\begin{table}
 \caption{The number of mesh points in 2D and 3D models}
 \label{tbl:mesh}
 \begin{center}
 \begin{tabular}{l|ccc}
         & points & faces & cells \\ \hline
  2D model & 27,062 & 53,130 & 13,200 \\
  3D model & 1,434,041 & 4,197,500 & 1,382,000
 \end{tabular}
 \end{center}
\end{table}

\subsection{Frequency of Helmholtz resonance}

The resonance frequency of a three dimensional Helmholtz
resonator \cite{Fletcher} is given by
\begin{equation}
 f_0 = \frac{1}{2\pi}\sqrt{\frac{K}{m}}
     = \frac{c}{2\pi}\sqrt{\frac{S}{VL}}
\end{equation}
where $S$ is its aperture size and $V$ is volume of its cavity.
In case of 3D, it is known that $L$ can be written as
$L \cong 2 \times 8a/3\pi$, where $a$ is the radius of the aperture.
As in consequence, we get $f_{3D}$ as
\begin{equation}
 f_0 \approx \frac{c}{8}\sqrt{\frac{3 \alpha}{V}} 
  \approx \frac{c}{2\pi}\sqrt{\frac{1.85a}{V}}.
\label{eq:f3d}
\end{equation}

Next, let's estimate $f_{3D}$ for our 3D model.
Parameters of the model are
\begin{equation}
\begin{split}
 V &= 9.425 \times 10^{-6} \\
 S &= 2.5 \times 10^{-5} \\
 a &= 2.82 \times 10^{-3} \\
 c &= 340.
\end{split} 
\end{equation}
Using these parameters, we can estimate the value of Eq.~(\ref{eq:f3d})
\begin{equation}
 f_{3D} \approx 1.273 \times 10^{3},
 \label{eq:f3dValue}
\end{equation}
which should be compared to values obtained by numerical calculations.

\begin{figure}
\begin{center}
\includegraphics[width=4.33cm]{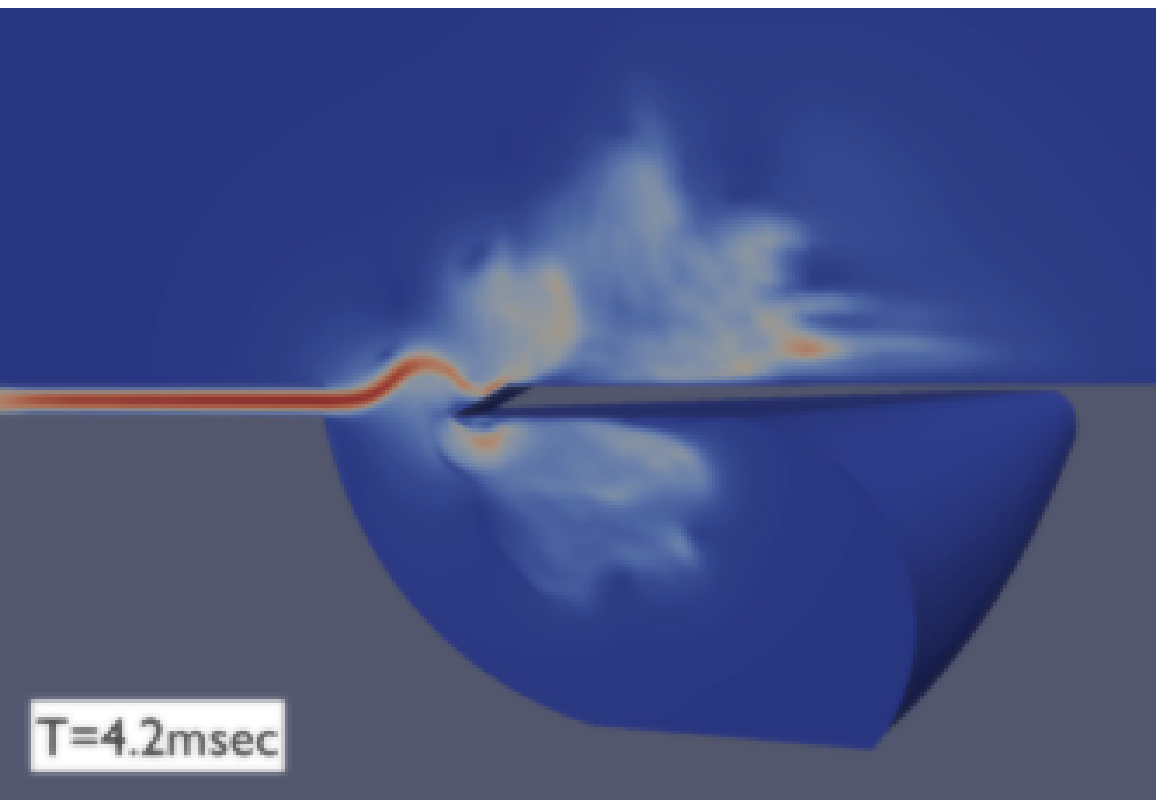}
\includegraphics[width=7cm]{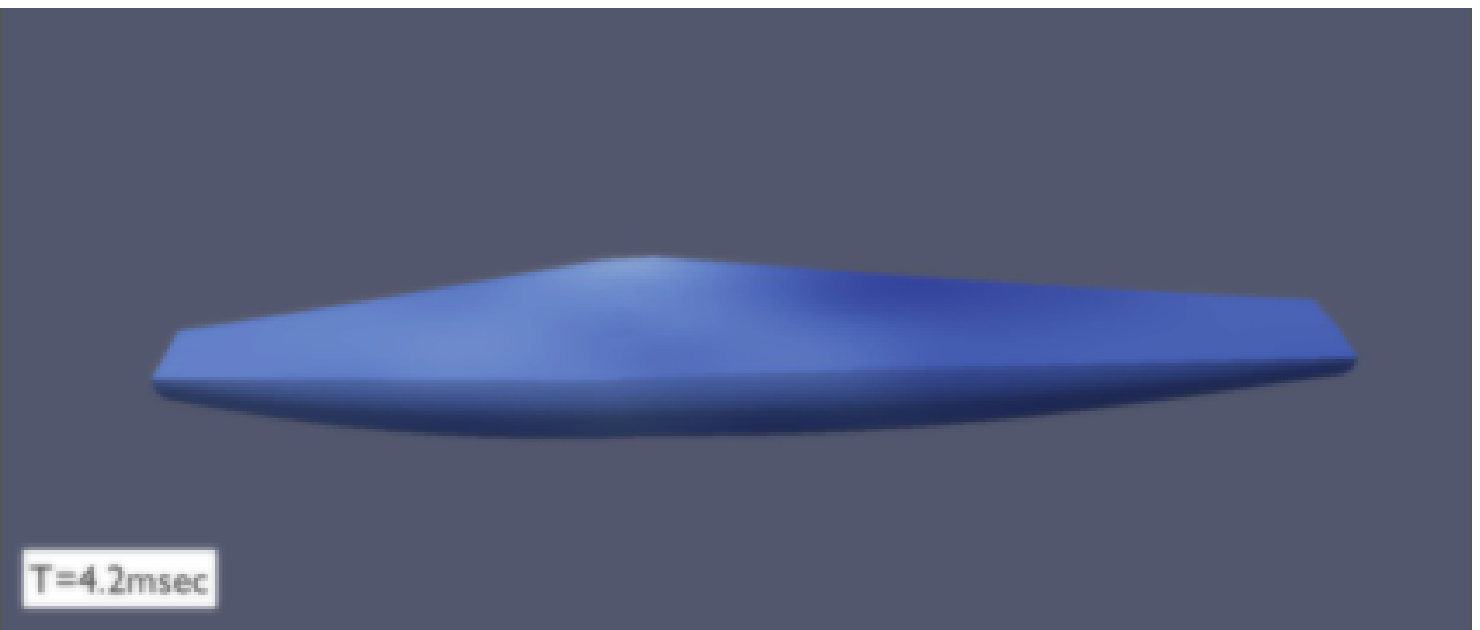}\\
\includegraphics[width=4.33cm]{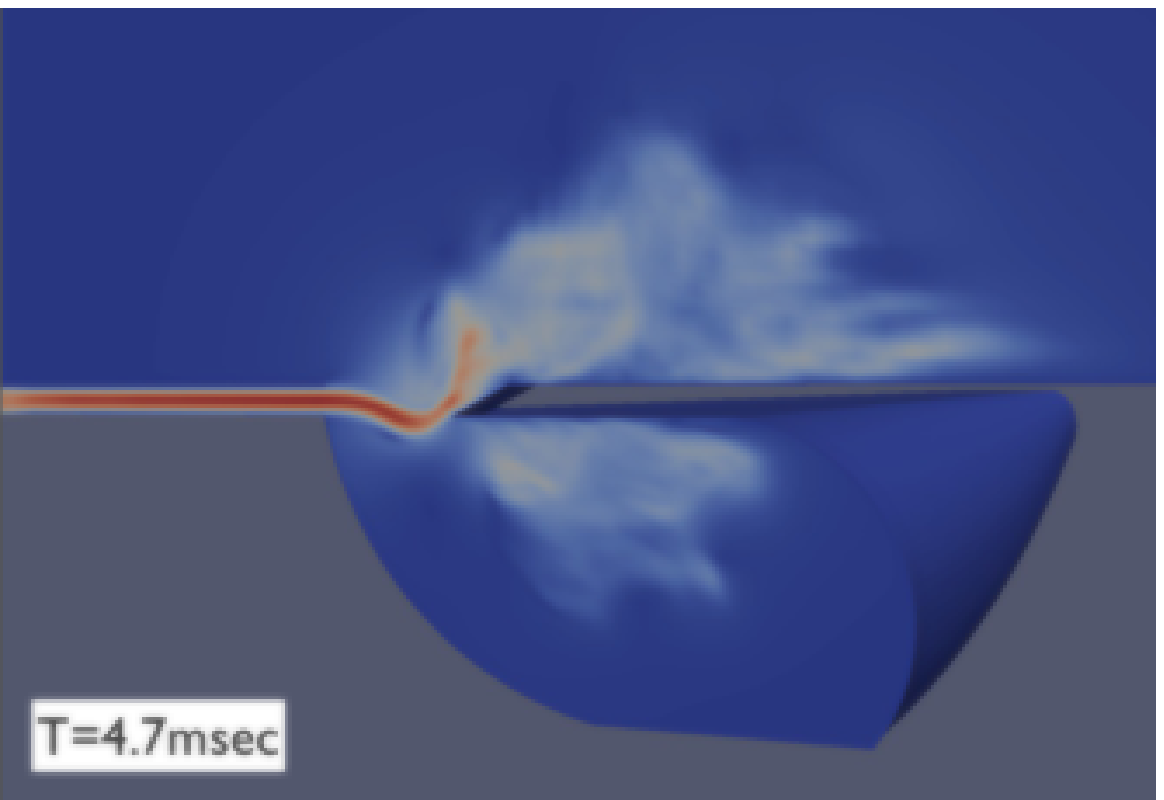}
\includegraphics[width=7cm]{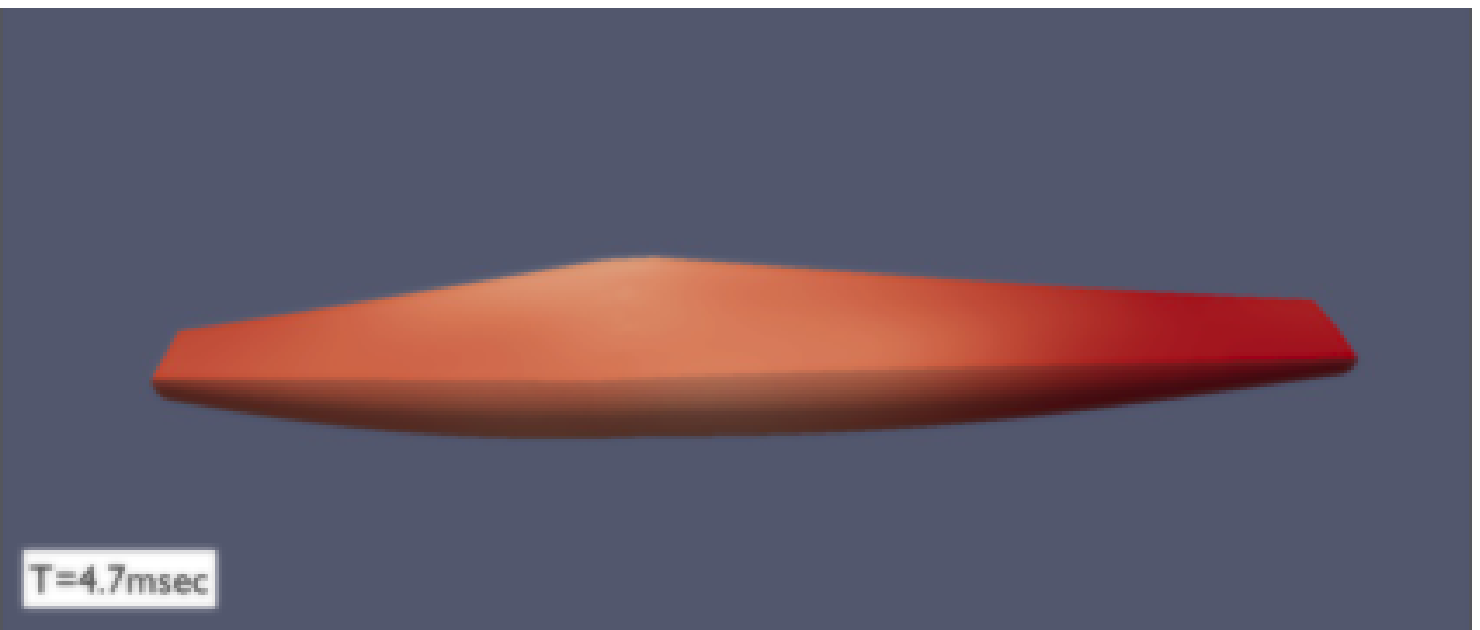}
\caption{\small Internal fields in the 3D ocarina cavity:
The left pictures represent snapshots of fluid velocity around an edge in
the 3D ocarina model at $t=4.2$ msec and $t=4.7$ msec,
which is shown by cutting at the center of the aperture;
the right figures are distribution of pressure values in the cavity
at $t=4.2$ msec and $t=4.7$ msec.}
\label{fig:3dSnapshot}
\end{center}
\end{figure}

\begin{figure}
\centering{
\includegraphics[width=6cm]{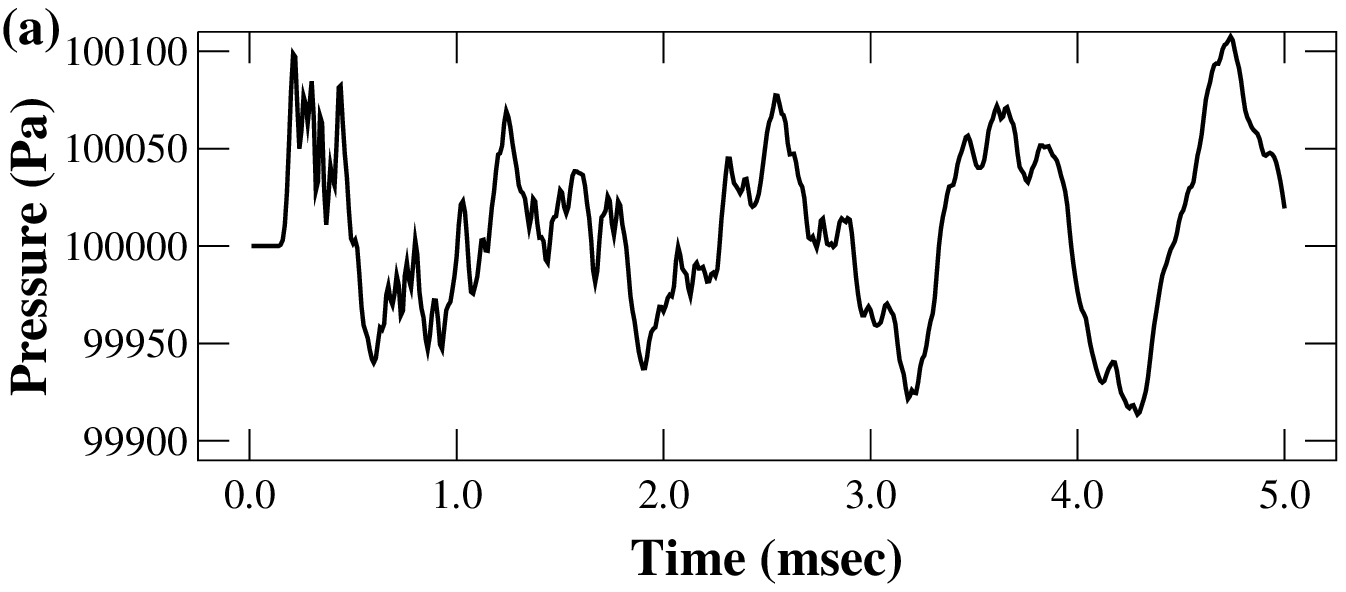}\hspace{5mm}
\includegraphics[width=6cm]{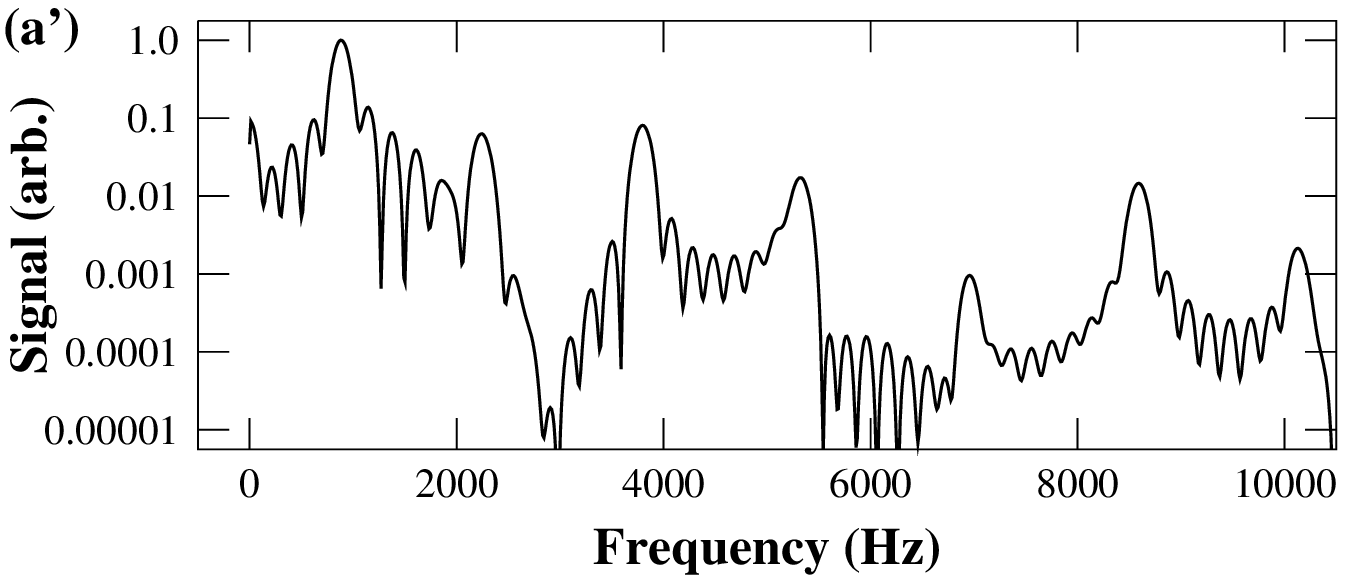}\\
\includegraphics[width=6cm]{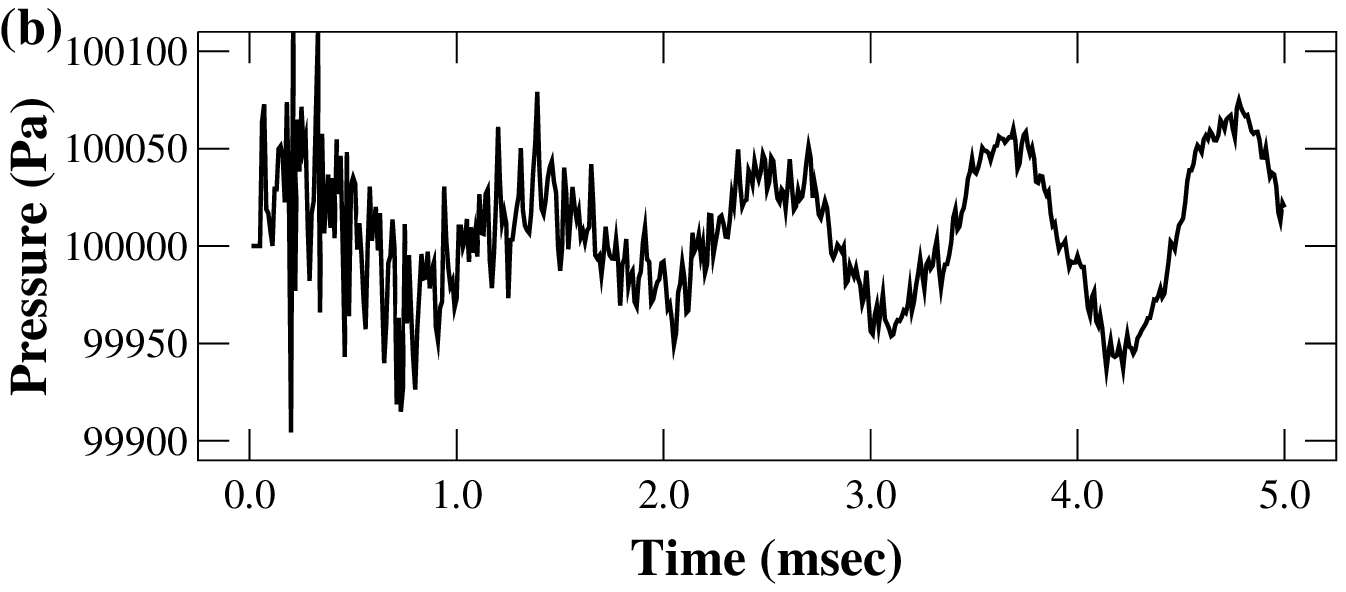}\hspace{5mm}
\includegraphics[width=6cm]{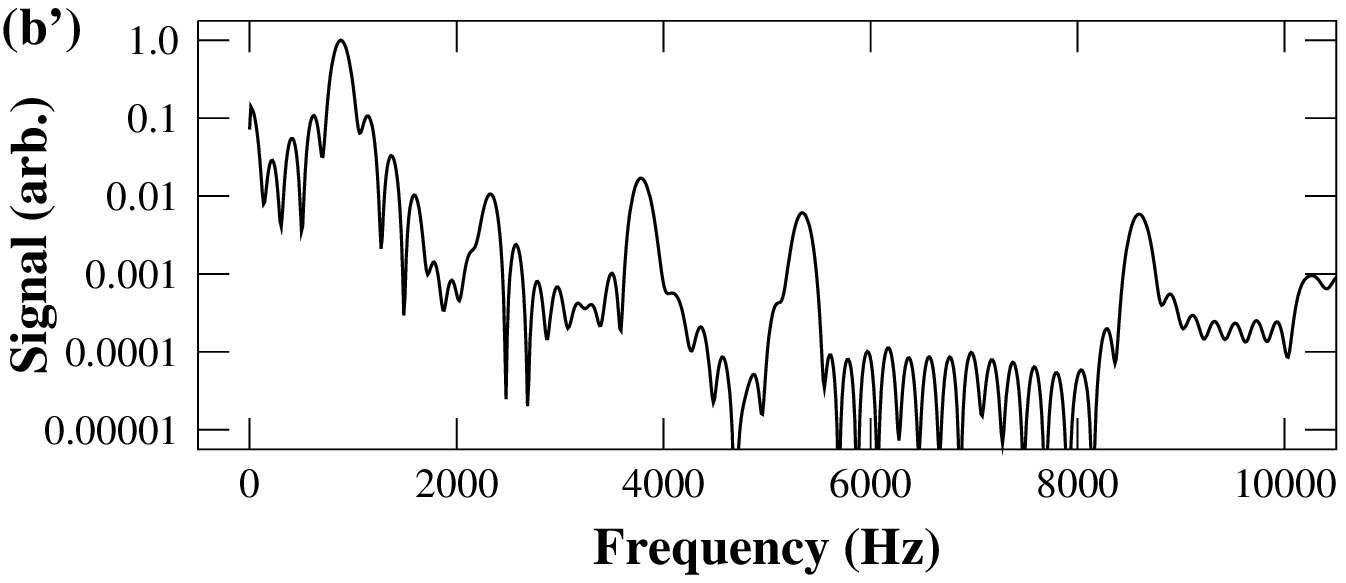}\\
\includegraphics[width=6cm]{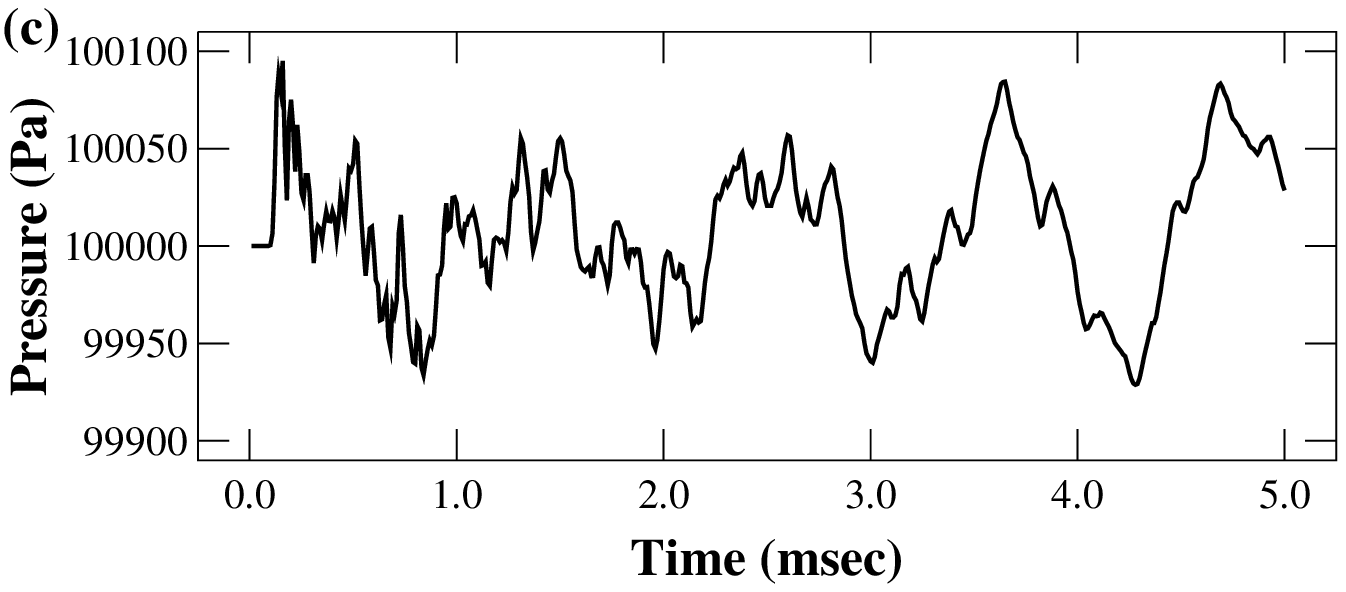}\hspace{5mm}
\includegraphics[width=6cm]{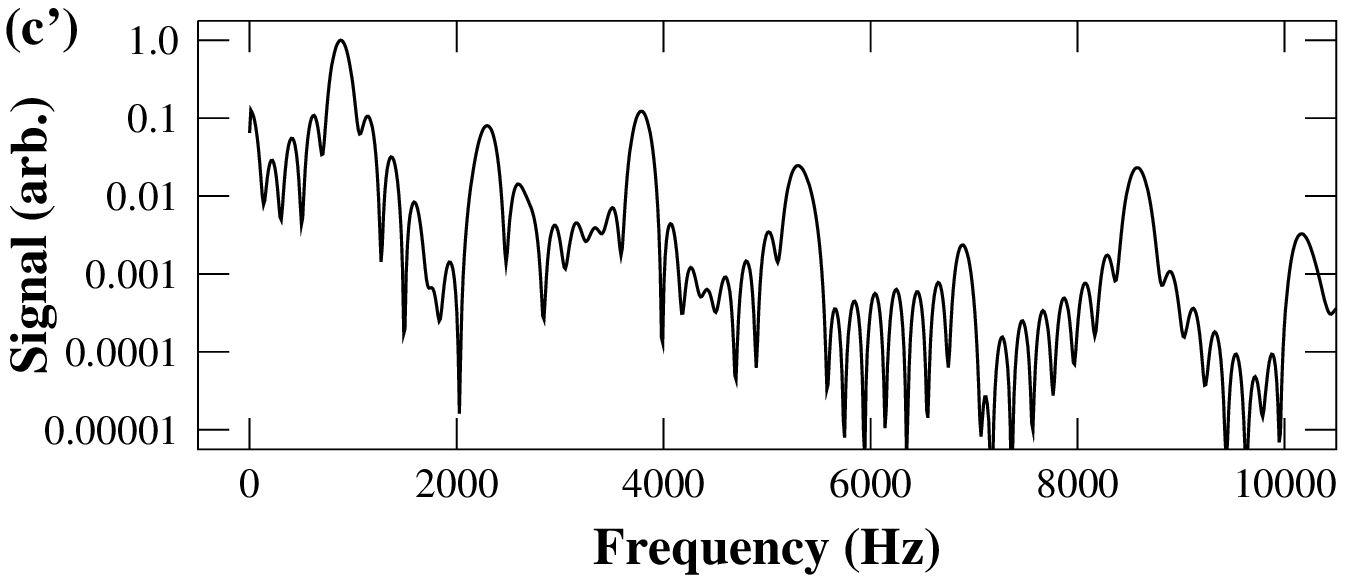}
}
\caption{\small Oscillation observed in 3D ocarina driven by a 10 m/s jet:
each figure shows pressure values detected
in the three observation points (a), (b), and (c), respectively.
 The corresponding normalized spectrum is shown in (a'), (b'), and
 (c').}
\label{fig:3dP}
\end{figure}

\subsection{Oscillation in ocarina}

The fluid velocity around the edge and the pressure in the cavity
obtained by the compressible LES solver
({\tt coodles} in OpenFOAM v1.5) are shown in Fig.\ref{fig:3dSnapshot},
which are snapshots at $t=4.2$ msec and $t=4.7$ msec.
Oscillations of sound pressure values detected at points
(a), (b), and (c) in Fig.\ref{fig:3dMesh} are shown in Fig.\ref{fig:3dP}
as well as Fourier transformed values.
It is seen that the oscillation is synchronized
over the whole cavity.

The frequency is determined as $8.8 \times 10^2$ Hz
from the Fourier transformation of the pressure data,
This is lower than expected values
from length resonances of the air-column,
whose frequencies are obtained as $1.7 \times 10^2$ Hz,
$1.4 \times 10^2$ Hz, and $2.1 \times 10^2$ Hz from 
$10$ cm/$\frac12\lambda$, $6$ cm/$\frac14\lambda$,
and $4$ cm/$\frac14\lambda$, respectively.
The expected frequency as the Helmholtz oscillator,
Eq.(\ref{eq:f3dValue}), gives still larger than 
the frequency observed here.
However, the theoretical calculation of Eq.(\ref{eq:f3dValue})
contains various uncertain factors on the open-end correction
around the edge hole.

We have an actual ocarina instrument which was used to design
the cavity for this calculation.  The size of the ocarina
is almost the same as the geometry given in this paper.
According to an instruction of the ocarina,
'A$_5$' (higher {\it la}) is the lowest note of this small instrument,
i.e., the note when all the tone holes are closed.
The frequency value of A$_5$ is assigned as 880 Hz
in a usual musical scale.  This is just the frequency observed.
Thus, when we consider the frequency observed, we can conclude that
our three dimensional model almost exactly reproduce the basic oscillation
of the ocarina.

If we assume the basic mechanism in an ocarina as Helmholtz oscillation,
it is expected that the internal oscillation of the cavity
does not contain higher harmonics.
However, the Fourier transformed spectrum in Fig.\ref{fig:3dP}
contains several peaks.
This is mainly because the oscillation observed has not been grown
sufficiently.  Five milli-second used to analyze the oscillation
is too short to observe stable oscillations in musical instruments.
Moreover, a slightly higher blow velocity might be appropriate
to drive the basic frequency.
It is expected to clarify these points by executing
more simulations with longer time.

\section{Conclusion}
\label{sec:conclusion}

In summary, we simulated an oscillation of a two and three dimensional
ocarina by the use of the compressible LES solver in OpenFOAM 1.5.
In both models, the resonance
showed characteristic properties of the Helmholtz resonance.
We conclude that the Helmholtz oscillation in a small air-reed
instrument is properly reproduced by the compressible LES calculations.

As future works, various interesting investigations can be planned
with the following theme:
simulations of tone holes in which dynamic transitions between
multiple frequencies;
transitions between the Helmholtz resonance and the column-length
resonance when we vary the geometry of the instrument;
relation between the type of oscillations and sounding mechanisms
\cite{Lighthill}.

\end{document}